\documentclass[11pt]{article}
\usepackage[T1]{fontenc}
\usepackage{lmodern}
\usepackage{geometry} 
\usepackage{graphicx}
\usepackage{amsmath,amssymb,amsthm,amsfonts,dsfont,mathtools}
\usepackage{relsize}
\usepackage[margin=1cm]{caption}
\usepackage{wrapfig}
\usepackage{frame,color}
\usepackage{environ,subfig}
\usepackage{xspace}
\usepackage{framed}
\usepackage{comment}
\usepackage{changepage}
\usepackage{enumerate}
\usepackage{algorithm}
\usepackage[noend]{algpseudocode}
\usepackage{algorithmicx}
\usepackage{mathrsfs}
\usepackage{soul}
\usepackage{authblk}
\usepackage{hyperref}
\usepackage[utf8]{inputenc}
\usepackage{tikz}
\usepackage{makecell}
\usepackage{microtype,hyperref}
   \hypersetup{%
      breaklinks,%
      ocgcolorlinks, colorlinks=true,%
      urlcolor=[rgb]{0.25,0.0,0.0},%
      linkcolor=[rgb]{0.5,0.0,0.0},%
      citecolor=[rgb]{0,0.2,0.445},%
      filecolor=[rgb]{0,0,0.4},
      anchorcolor=[rgb]={0.0,0.1,0.2}%
   }
\usepackage{url}
\usepackage{breakurl}

\usepackage{lineno}

\newcommand{\PAPER}[1]{}

\algnewcommand{\Inputs}[1]{%
  \State \textbf{Inputs:}
  \Statex \hspace*{\algorithmicindent}\parbox[t]{.8\linewidth}{\raggedright #1}
}
\algnewcommand{\Initialize}[1]{%
  \State \textbf{Initialize:}
  \Statex \hspace*{\algorithmicindent}\parbox[t]{.8\linewidth}{\raggedright #1}
}
\algnewcommand{\TurnOne}[1]{%
  \State \textbf{Timestep 1:}
  \Statex \hspace*{\algorithmicindent}\parbox[t]{.8\linewidth}{\raggedright #1}
}

\makeatletter
 {\par\unskip\endMakeFramed}
\makeatother

\makeatletter
\newsavebox{\@brx}
\newcommand{\llangle}[1][]{\savebox{\@brx}{\(\m@th{#1\langle}\)}%
  \mathopen{\copy\@brx\mkern2mu\kern-0.9\wd\@brx\usebox{\@brx}}}
\newcommand{\rrangle}[1][]{\savebox{\@brx}{\(\m@th{#1\rangle}\)}%
  \mathclose{\copy\@brx\mkern2mu\kern-0.9\wd\@brx\usebox{\@brx}}}
\makeatother

\NewEnviron{fcodebox}[1]%
{\fbox{%
\begin{minipage}{#1\linewidth}
\begin{codebox}
\BODY
\end{codebox}
\end{minipage}
}}

\newtheorem{fact}{Fact}

\newtheorem{theorem}[fact]{Theorem}
\newtheorem{lemma}[fact]{Lemma}

\newtheorem{problem}[fact]{Problem}

\newcommand{\ignore}[1]{}

\newcommand{\OPT}{\textrm{OPT}}
\newcommand{\tO}{\widetilde{O}}

\newcommand{\polylog}{\mathop{\mathrm{polylog}}}

\newcommand{\Prob}{\operatorname{Pr}}

\newcommand{\floor}[1]{\left\lfloor #1 \right\rfloor}

\newcommand{\pmax}{p_\mathrm{max}}
\newcommand{\wmax}{w_\mathrm{max}}
\newcommand{\Ip}{\mathcal{I}^+}
\newcommand{\In}{\mathcal{I}^-}
\newcommand{\Igreedy}{\widetilde{\mathcal{I}}}
\newcommand{\Iopt}{\mathcal{I}^*}
\newcommand{\Idiff}{\mathcal{I}_{\Delta}}
\newcommand{\Sp}{S^+}
\newcommand{\Sn}{S^-}
\newcommand{\Ss}{\mathcal{S}^*}
\newcommand{\DK}{\textsc{DiffKnapsack}}

\makeatletter
\DeclareRobustCommand\onedot{\futurelet\@let@token\@onedot}
\def\@onedot{\ifx\@let@token.\else.\null\fi\xspace}

\makeatother

\title{Simple and Faster Algorithms for Knapsack}

\author[1]{Qizheng He\thanks{qizheng6@illinois.edu.}}
\author[2]{Zhean Xu\thanks{xuza21@mails.tsinghua.edu.cn.}}
\affil[1]{Department of Computer Science, University of Illinois at Urbana-Champaign}
\affil[2]{Department of Computer Science and Technology, Tsinghua University}

\begin{document}
\date{}
\maketitle
\thispagestyle{empty}

\begin{abstract}
In this paper, we obtain a number of new simple pseudo-polynomial time algorithms on the well-known \emph{knapsack} problem, focusing on the running time dependency on the number of items $n$, the maximum item weight $\wmax$, and the maximum item profit $\pmax$. Our results include:
\begin{itemize}
\item An $\tO(n^{3/2}\cdot \min\{\wmax,\pmax\})$-time randomized algorithm for \emph{0-1 knapsack}, improving the previous $\tO(\min\{n\wmax\pmax^{2/3},n\pmax\wmax^{2/3}\})$ [Bringmann and Cassis, ESA'23] for the small $n$ case.
\item An $\tO(n+\min\{\wmax,\pmax\}^{5/2})$-time randomized algorithm for \emph{bounded knapsack}, improving the previous $O(n+\min\{\wmax^3,\pmax^3\})$ [Polak, Rohwedder and Wegrzyck, ICALP'21].
\end{itemize}

\end{abstract}


\section{Introduction}

In the \emph{0-1 knapsack} problem, we are given a knapsack capacity $W\in\mathbb{Z}^+$ and a set $\mathcal{I}$ of $n$ items $(w_1,p_1),\cdots,(w_n,p_n)$, where item $i$ has a \emph{weight} $w_i\in \mathbb{Z}^+$ and a \emph{profit} $p_i\in \mathbb{Z}^+$. Our goal is to select a subset $S\subseteq [n]$ of items, so as to maximize the total profit $\OPT=\sum_{i\in S}p_i$ subject to the constraint $\sum_{i\in S} w_i\le W$. In the \emph{bounded knapsack} problem, each item $i$ is additionally given a multiplicity $m_i$, indicating it can be selected up to $m_i$ times. Due to its simplicity, the knapsack problem is commonly used as teaching material in undergraduate algorithm courses.

Knapsack is known to be NP-hard, but when the input numbers are small, there is a standard $O(nW)$ dynamic programming algorithm \cite{bellman1958dynamic}. A series of work has studied the pseudo-polynomial time algorithms with respect to various parameters, including $n$, $W$, $\OPT$, the maximum weight $\wmax$ and the maximum profit $\pmax$.
The known algorithms for 0-1 knapsack and bounded knapsack are summarized in Table \ref{table1}. We can, without loss of generality, assume $\wmax\leq W< \sum_{i=1}^{n}w_i\leq n\cdot \wmax$, and $\pmax\leq \OPT < \sum_{i=1}^{n}p_i \leq n\cdot \pmax$.

According to the $\Omega((n+W)^{2-\delta})$-time conditional lower bound \cite{DBLP:journals/talg/CyganMWW19,DBLP:conf/icalp/KunnemannPS17}, the $O(nW)$ time complexity is basically optimal if we parameterize by $W$. However, since $W$ can be as large as $\Theta(n\wmax)$, which gives $O(n^2\wmax)$ time complexity, there might be faster algorithms when $\wmax$ is significantly smaller than $W$. Thus, recent works prefer to focus on parameterizing the running time in terms of $n$, $\wmax$ and $\pmax$.

Notably, Polak et al.~\cite{DBLP:conf/icalp/0001RW21} presented an $O(n+\min\{\wmax^3,\pmax^3\})$-time algorithm, which is the first algorithm that has only linear dependency on the number of items $n$.
The $\tO(n\wmax\pmax^{2/3})$\footnote{The $\tO$ notation hides polylogarithmic factors.}-time algorithm by Bringmann and Cassis~\cite{DBLP:journals/corr/abs-2305-01593} is the first to break the cubic barrier $O(n^3)$ in the regime $\pmax\approx \wmax \approx n$.

\begin{table}[!htbp]\centering
\begin{tabular}{|c|c|c|}
\hline
Running time & References & Works for bounded?\\
\hline
$O(n+\min\{\OPT^2,W^2\})$ & \cite{bellman1958dynamic} & \\
\hline
$O(n\cdot \min\{W,\OPT\})$ & \cite{bellman1958dynamic} & \checkmark\\
\hline
$O(n\pmax\wmax)$ & \cite{DBLP:journals/jal/Pisinger99} &  \\
\hline
$O(n^2\wmax)$ & \cite{DBLP:journals/jal/Pisinger99} &  \\
\hline
$O(n^2\pmax)$ & \cite{DBLP:journals/jco/KellererP04,DBLP:books/daglib/0010031} &  \\
\hline
$\tO(n+\wmax\cdot W)$ & \cite{DBLP:journals/jco/KellererP04,DBLP:conf/stoc/BateniHSS18,DBLP:conf/icalp/AxiotisT19} & \checkmark\\
\hline
$\tO(n+\pmax\cdot W)$ & \cite{DBLP:conf/stoc/BateniHSS18} & \checkmark\\
\hline
$\tO(n\cdot \min\{\wmax^2,\pmax^2\})$ & \cite{DBLP:conf/icalp/AxiotisT19} & \\
\hline
$\tO(n\wmax^2)$ & \cite{DBLP:journals/talg/EisenbrandW20} & \checkmark\\
\hline
$O(n+\min\{\wmax^3,\pmax^3\})$ & \cite{DBLP:conf/icalp/0001RW21} & \checkmark\\
\hline
$\tO(n+(W+\OPT)^{3/2})$ & \cite{DBLP:conf/icalp/BringmannC22} &  \\
\hline
$\tO(n\wmax\pmax^{2/3})$
& \cite{DBLP:journals/corr/abs-2305-01593} & \\
\hline
$\tO(n\pmax\wmax^{2/3})$ & \cite{DBLP:journals/corr/abs-2305-01593} & \\
\hline
$\tO(n^{3/2}\cdot \min\{\wmax,\pmax\})$ & \textbf{new} & \\
\hline
$\tO(n+\min\{\wmax,\pmax\}^{5/2})$ & \textbf{new} & \checkmark\\ 
\hline
\end{tabular}
\caption{Pseudopolynomial-time algorithms for 0-1 knapsack and bounded knapsack. Here $n$ denote the number of items, $W$ denote the weight budget, $\wmax$ denote the largest weight of any item, $\pmax$ denote the largest profit of any item, and $\OPT$ denote the optimal profit.
}
\label{table1}
\end{table}

\paragraph{Remark on concurrent independent works.}

We briefly summarize a number of concurrent works that appeared in the past month. Jin~\cite{jin2023solving} designed an algorithm for 0-1 knapsack with running time $\tO(n+\wmax^{5/2})$. Chen et al.~\cite{chen2023faster} presented a slightly faster algorithm for bounded knapsack that runs in time $\tO(n+\wmax^{12/5})$. Based on their results, Jin~\cite{jin202301} further improved his algorithm to $\tO(n+\wmax^2)$ for 0-1 knapsack, and concurrently Bringmann~\cite{bringmann2023knapsack} also obtained $\tO(n+\wmax^2)$-time for bounded knapsack. Since there is an $\Omega((n+\wmax)^{2-\delta})$-time conditional lower bound for 0-1 knapsack~\cite{DBLP:journals/talg/CyganMWW19,DBLP:conf/icalp/KunnemannPS17}, the latter two results are basically optimal.

We highlight two key differences between our results with these works: 1)~All these works focus on improving the running time dependency solely on $\wmax$ (apart that Jin~\cite{jin2023solving} also had an alternative $\tO(n\wmax^{3/2})$-time algorithm), while our first $\tO(n^{3/2} \wmax)$-time algorithm is not only simple, but also faster when $n$ is small. 2)~The techniques of all these papers are based on a theorem in additive combinatorics by Galil and Margalit~\cite{DBLP:journals/siamcomp/GalilM91} for dense subset-sum (which was recently refined by Bringmann and Wellnitz~\cite{DBLP:conf/soda/BringmannW21}), while our second algorithm uses a number-theoretical theorem by Erd\H{o}s and Graham~\cite{erdos1972linear} (whose proof depends on Kneser's theorem~\cite{kneser1953abschatzung} in additive combinatorics), which is a very different path from the others. Although our running time dependency on $\wmax$ doesn't seem to be advantageous, the algorithm is quite simple, and the way for extending Erd\H{o}s--Graham in our correctness proof might be interesting in its own right.


\ignore{
\begin{table}[!htbp]\centering
\begin{tabular}{|c|c|c|}
\hline
Running time & Reference & Best?\\
\hline
$O(n\cdot W)$ & \cite{bellman1958dynamic} & \\
\hline
$\tO(n+\wmax\cdot W)$ & \cite{DBLP:journals/jco/KellererP04,DBLP:conf/stoc/BateniHSS18,DBLP:conf/icalp/AxiotisT19} & \\
\hline
$\tO(n+\pmax\cdot W)$ & \cite{DBLP:conf/stoc/BateniHSS18} & \\
\hline
$\tO(n\wmax^2\min\{n,\wmax\})$ & \cite{DBLP:conf/stoc/BateniHSS18} & $\times$\\
\hline
$\tO(n\wmax^2)$ & \cite{DBLP:journals/talg/EisenbrandW20} & \\
\hline
$O(n+\min\{\wmax^3,\pmax^3\})$ & \cite{DBLP:conf/icalp/0001RW21} & \checkmark \\
\hline
$\tO(n+\min\{\wmax,\pmax\}^{5/2})$ & new? & \checkmark \\ 
\hline
\end{tabular}
\caption{Pseudopolynomial-time algorithms for bounded Knapsack.}
\end{table}
}

\paragraph{The proximity results.}

Assume the items are sorted by \emph{efficiency}, s.t.\ $p_1/w_1\ge \cdots \geq p_n/w_n$ (for bounded knapsack, we create $m_i$ copies of the $i$-th item and formalize as 0-1 knapsack). Intuitively, the higher efficiency items are preferable to the lower ones. Thus, the idea is to greedily take the \emph{maximal prefix solution} $\Igreedy=\{(p_1,w_1),\dots,(p_k,w_k)\}$, where $k=\max\{k\mid w_1+\cdots+w_k\le W\}$, and then focus on its difference from the optimal solution $\Iopt$. Polak et al.~\cite{DBLP:conf/icalp/0001RW21} proved that there exists an optimal solution $\Iopt$, such that the symmetric difference $\Delta(\Igreedy,\Iopt)$ between $\Iopt$ and the greedy solution $\Igreedy$ is small.
\begin{lemma}\label{lemma:proximity}
There exists an optimal solution $\Iopt$ s.t.\ $|\Delta(\Igreedy,\Iopt)|=O(\wmax)$.
\end{lemma}

Let $\Ip = \{(w_{k+1},p_{k+1}),\cdots,(w_{n},p_{n})\}$ be the set of items not taken by the greedy solution $\Igreedy$, $\In = \{(w_{1},p_{1}),\cdots,(w_{k},p_{k})\}$ be the set of items taken by $\Igreedy$, and $W'=W-\sum_{i=1}^{k}w_i$. The original bounded knapsack problem can be reduced to the following 0-1 knapsack problem (with negative items) in $\tO(n)$ time \cite{DBLP:conf/icalp/0001RW21}:
\begin{problem}[\DK]
    Select $\Sp\subseteq\{k+1,\dots,n\}$ and $\Sn\subseteq\{1,\dots,k\}$ such that $\sum_{i\in\Sp} p_i - \sum_{i\in\Sn} p_i$ is maximized subject to the constraint $\sum_{i\in\Sp} w_i - \sum_{i\in\Sn} w_i \le W'$, where $W'\in[0,\wmax)$.
\end{problem}

\paragraph{Overview of our results.}

In this work, we mainly focus on designing faster algorithms for knapsack with running time depending on the parameters $n$ and $\wmax$:
\begin{itemize}
\item The first result is an extremely simple $\tO(n^{1.5}\wmax)$-time randomized algorithm for 0-1 knapsack, which is the best known in the small $n$ regime (i.e., $n=\tO(\wmax^{2/3})$). We notice that a simple \emph{random permutation} technique is helpful for solving 0-1 knapsack. Applying this technique to the standard $O(n^2\wmax)$ dynamic programming algorithm results in an $\tO(\sqrt n)$ factor speedup.

\item The second result is a simple $\tO(n+\wmax^{2.5})$-time randomized algorithm for bounded knapsack. The result is obtained by combining our first algorithm with a novel structural theorem on the difference between the greedy solution and an optimal solution, which is proved via an extension of Erd\H{o}s--Graham (the proof does not seem to be trivial, but the algorithm itself is simple).

\end{itemize}
Combing both of our results, we obtain a strict improvement over the $\tO(n\wmax\pmax^{2/3})$-time algorithm by Bringmann and Cassis~\cite{DBLP:journals/corr/abs-2305-01593} in all regimes. 

\section{0-1 Knapsack in $\tO(n^{3/2}\wmax)$ Time}\label{sec:random_permutation}

In this section, we present an $\tO(n^{3/2}\wmax)$-time algorithm for 0-1 knapsack, using a simple random permutation technique. Let $W_{\mathrm{opt}}$ denote the total weight of the items in an optimal solution $\Iopt$. The intuition is that after randomly permuting the items, if we consider any prefix of the items $(w_1,p_1),\dots,(w_i,p_i)$, the total weight $W_{\mathrm{opt}}^{(i)}$ of the items in $\Iopt$ within this prefix should concentrate towards the expectation $\mu_i\triangleq \mathbb{E}[W_{\mathrm{opt}}^{(i)}]=\frac{i}{n}\cdot W_{\mathrm{opt}}$, with deviation bounded by $\Delta_i\triangleq \tO(\sqrt{i}\cdot \wmax)$ w.h.p. Easy to see that $W-\wmax<W_{\mathrm{opt}}\leq W$, since otherwise one can always add more items to the solution without exceeding the weight limit $W$, unless one has used up all the items. Therefore the (unknown) expectation of $W_{\mathrm{opt}}^{(i)}$ can also be narrowed within a short interval of length $\wmax$.

The advantage of this concentration phenomena is that it does not require any knowledge on the (unknown) optimal solution. Therefore, if we perform the standard dynamic programming algorithm (where $f[i][j]$ denote the maximum profit one can get, using a subset of items with total weight $j$ among the first $i$ items) and consider the items one by one, it suffices to compute $f[i][j]$ for a short interval $J_i\triangleq [\mu_i-\Delta_i,\mu_i+\Delta_i]$ of $j$, using the DP formula. This interval will contain the total weight $W_{\mathrm{opt}}^{(i)}$ of the optimal solution truncated by the prefix 1$\sim i$ w.h.p., and if $f[i-1][W_{\mathrm{opt}}^{(i-1)}]$ is computed correctly, so will be $f[i][W_{\mathrm{opt}}^{(i)}]$. (Other DP entries $f[i][j]$ for $j\in J_i$ may be incorrectly small, but we only care about the entry $f[i][W_{\mathrm{opt}}^{(i)}]$.)

In this way, we obtain an algorithm with running time $\tO(n\cdot \sum_{i=1}^n \Delta_i)=\tO(n\cdot \sqrt{n}\wmax)=\tO(n^{3/2}\wmax)$. We emphasize that this algorithm is extremely simple, and can be implemented in under 10 lines of code: see Alg.~\ref{alg:rand_knapsack}.

\begin{algorithm}[!htbp]
\caption{Algorithm for 0-1 knapsack}
\label{alg:rand_knapsack}
\begin{algorithmic}[1]
\State Randomly permute $(w_1,p_1),\dots,(w_n,p_n)$.
\State Initialize $f[\cdot ][j]=0$ for $j\geq 0$, and $f[\cdot ][j]=-\infty$ for $j<0$. 
\For {$i=1,\dots,n$}
    \State Set $\mu_i=\frac{i}{n}\cdot W$, and $\Delta_i=\tO(\sqrt{i}\cdot \wmax)$.
    \For {$j=\mu_i-\Delta_i,\dots,\mu_i+\Delta_i$}
        \State $f[i][j]=\max\{f[i-1][j],f[i-1][j-w_i]+p_i\}$.
    \EndFor
\EndFor
\State Return $\max_{0\leq j\leq W} f[n][j]$.
\end{algorithmic}
\end{algorithm}

We remark that this technique is not new; e.g., it has been used in the context of subset sum and discrepency theory. However, to the best of our knowledge, no published paper has applied this technique directly to the knapsack problem\footnote{Though the application of this idea on knapsack has been repeatedly rediscovered in the competitive programming community: in 2020, the second author has used this idea to create a programming contest problem \url{http://acm.hdu.edu.cn/showproblem.php?pid=6804}, and similar ideas appeared even earlier in 2016 \url{https://codeforces.com/blog/entry/50036}.}, and surprisingly, this simple idea yields an improvement on the current best algorithms in the small $n$ regime.

In the following we more formally present the analysis of this idea. The probabilistic analysis of the random permutation technique is based on Hoeffding's inequality in the situation of sampling without replacement~\cite{Hoeffding}.

\begin{lemma}[Hoeffding's inequality]
     Given a set $\mathcal{A}$ consists of $n$ values $a_1,a_2,\cdots,a_n$. Let $\mu=\bigl(\sum_{j=1}^{n}a_j\bigr)/n$, and let $X_1,X_2,\cdots,X_i$ denote a random sample without replacement from $\mathcal{A}$. If $\ell\le a_i\le r$ for all $i$, then for any $t>0$ and $i\geq 1$,
    \[
        \Prob\left[\biggl|\sum_{j=1}^{i}X_j - \mu i\biggr| \ge ti\right]
        \leq
        2\exp\left(-\frac{2it^2}{(r-\ell)^2}\right).
    \]
\end{lemma}


%
\noindent Notice that any prefix of the random permutation is a random sample without replacement from $\mathcal{A}$. Let $a_i=w_i$ if the $i$-th item is in the optimal solution, otherwise $a_i=0$, so that $\sum_{j=1}^i X_j=W_{\mathrm{opt}}^{(i)}$, and $\mu=W_{\mathrm{opt}}/n$. Set $\ell=0$, $r=\wmax$ and $t=\sqrt{\log n}\cdot \wmax/\sqrt{i}$, the claimed result holds w.h.p.\ via a union bound on $i$.
We thus obtain: 
\begin{theorem}\label{thm:01_knapsack_random}
    There is an $\tO(n^{3/2}\wmax)$-time randomized algorithm for 0-1 Knapsack that is correct w.h.p.
\end{theorem}
\noindent By known techniques~\cite{DBLP:conf/icalp/0001RW21}, we can replace $\wmax$ with $\pmax$ in the running time of Theorem~\ref{thm:01_knapsack_random}.
\begin{theorem}\label{thm:01_knapsack_random_pmax}
    There is an $\tO(n^{3/2}\pmax)$-time randomized algorithm for 0-1 Knapsack that is correct w.h.p.
\end{theorem}

\section{Bounded Knapsack in $\tO(n+\wmax^{5/2})$ Time}\label{sec:w_max}
It is also intriguing to study the complexity of knapsack for the small items case, when the maximum item weight $\wmax$ is much smaller compared to the number of items $n$. In this section, we focus on algorithms for bounded knapsack with running time only depending on $\wmax$ (apart from the linear dependency on $n$), and present an algorithm with $\tO(n+\wmax^{5/2})$ running time.


The proximity result in Lemma~\ref{lemma:proximity} already showed that the greedy solution $\Igreedy$ and an optimal solution $\Iopt$ cannot differ in more than $O(\wmax)$ items. However, it is \emph{unknown} that which items can be contained in this difference set $\Delta(\Igreedy,\Iopt)$. Therefore, in the previous $O(\wmax^3)$-time algorithm by Polak et al.~\cite{DBLP:conf/icalp/0001RW21}, they need to keep $O(\wmax)$ items for each of the $\wmax$ distinct weights, in total keeping $O(\wmax^2)$ items, and then solve the \DK~problem. Our key idea is to apply tools from the area of \emph{additive combinatorics} and \emph{number theory}, obtaining a structural theorem on the set of items that can appear in the difference set, thus reducing the number of items that we need to consider to only $\tO(\wmax)$.

\paragraph{A structural result for reducing the number of items.}
Our structural theorem for the difference set $\Delta(\Igreedy,\Iopt)$ is as follows:
\begin{theorem}\label{thm:item_reduction}
For the bounded knapsack problem with integer item weights bounded by $\wmax$, one can compute a superset $\Idiff$ of the symmetric difference $\Delta(\Igreedy,\Iopt)$ between the greedy solution and an optimal solution in $\tO(n+\wmax)$ time, such that $\Idiff$ only contains $\tO(\wmax)$ items.
\end{theorem}

Before proving this result, we first revisit a lemma by Erd\H{o}s and Graham in 1972~\cite{erdos1972linear}, on the Frobenius problem.

\begin{lemma}\label{lemma:erdos_graham}
\emph{(Erd\H{o}s--Graham)}
Given positive integers $v_1>\dots>v_k\ (k\ge 2)$ with $\gcd(v_1,\dots,v_k)=d$ (we call them the bases), any integer greater than $2\floor{\frac{v_1}{dk}}v_2-v_1$ and is divisible by $d$ can be expressed as a nonnegative integer linear combination of $v_1,\dots,v_k$.
\end{lemma}
This lemma was utilized by Chan and He~\cite{chan2020more}, for designing faster algorithms on the change-making problem and the unbounded knapsack problem. The high level idea is to first sort the items in decreasing order of $p_i/w_i$, then use the weight of the first $k$ items as the bases. In this way, if the total weight of the subsequent items used exceeds $2\wmax^2/k$, then the solution cannot be optimal, since we can replace these items with a nonnegative integer linear combination of the first $k$ items, obtaining a better profit while keeping the same total weight.

Our idea is to prove Theorem~\ref{thm:item_reduction} similarly using an extension of Erd\H{o}s--Graham. However, there are a number of difficulties when trying to extend its application from unbounded knapsack to bounded knapsack. First, we no longer have unlimited supply for the bases, since the multiplicity of each type of item is bounded. This makes some linear combinations unachievable. Secondly, the optimal solution might use items in the bases (when we choose bases from $\mathcal{I}\backslash \Igreedy$), making some of the bases unavailable. The appearance of these difficulties is not surprising, as the unbounded knapsack problem is generally considered as harder than the 0-1 (or bounded) knapsack problem~\cite{DBLP:conf/icalp/0001RW21,jin202301}, and most techniques designed for unbounded does not generalize. However, in the following proof, we show that all these issues can in fact be handled.

\begin{proof}[Proof for Theorem~\ref{thm:item_reduction}]
We consider the set of all items $S$ in $\Ip$ (resp., $\In$) that have similar weights in $[w,2w)$ as a batch, where $w=2^{i_0}$ for $i_0=0,\dots,\log \wmax$. Sort the items in $S$ in decreasing (resp., increasing) order of $p_i/w_i$. We try to prove that for each $r=2^{i_1}$ ($0\leq i_1\leq \log\wmax$), it suffices to keep only $O(r)$ items for each weight, except for $\tO(\wmax/r)$ weights.

We process the items one by one in the sorted order. Whenever we have seen $2r+2$ items with the same weight $w_i$, add $w_i$ to the set of bases $B$, until we have collected $9k$ initial bases, where $k=\wmax/r$. From the proximity result (Lemma~\ref{lemma:proximity}), the (unknown) optimal solution can only use (resp., discard) $\wmax$ items in $S$. The usage of these items by the optimal solution makes some of the bases unavailable (as in our proof, we want to replace items not in the bases and in the current solution, by items in the bases but not in the current solution, to obtain a better profit). However, it is guaranteed that there are still at least $8k$ bases in $B$, each of them has at least $r+2$ items appeared earlier that are not included in the optimal solution (i.e., at most $k$ bases are destroyed by the optimal solution). In this way, we aim to get a ``more robust'' version of Erd\H{o}s--Graham.

We call a subset $B'$ of $B$ as a $k$-subset, if $|B'|=|B|-k$. We call the greatest common divisor (gcd) of all elements of a $k$-subset as a $k$-gcd. If any of the $k$-gcds contains $p_i^{q_i}$ as a factor for some prime $p_i$ with corresponding exponent $q_i>0$, then $p_i$ must have been appeared as a factor for at least $|B|-k=\Omega(k)$ elements in $B$. The elements in $B$ have $\tO(k)$ prime factors in total (counting multiplicities), so we only need to consider $\tO(1)$ such primes $p_i$.

After collecting the initial bases, for the remaining items, continue processing them one by one. Whenever we have seen $2r+2$ items with the same weight $w_i$, we check whether $w_i$ can be divided by all current $k$-gcds. If true, then by Erd\H{o}s--Graham, the smallest multiple $m$ of $w_i$ exceeding $2(2w)^2/(8k)=w^2/k$ can always be expressed as a nonnegative integer linear combination of the remaining bases, after the optimal solution destroyed some of them. Furthermore, this linear combination is achievable, since $m\leq w^2/k+w_i< w(r+2)$, and each basis has at least $r+2$ available items, whose sum is at least $w(r+2)$. This implies that only the next $\frac{m}{w_i}-1\leq r$ items with weight $w_i$ can possibly appear in the optimal solution. Otherwise if false, then we add $w_i$ to the set $B$ of bases.

We next bound the total number of bases in $B$ after processing all items in the whole batch $S$. If a new basis $w_i$ is added, then there must exist a $k$-gcd that cannot divide $w_i$, which implies there is a $p_j^{q_j}$ that divides a $k$-gcd but does not divide $w_i$. One remaining issue is there are exponentially many $k$-subsets to check; therefore, we slightly loose the check condition. For each prime $p_j$, we keep track of a (possibly not tight) upper bound $q_j$ on the exponent of $p_j$ within the prime factorization among all $k$-gcds. Whenever we have collected at least $k+1$ new bases whose exponent of $p_j$ in the prime factorization is less than $q_j$ (call this a round), then we can safely decrease $q_j$ to a smaller value: since the optimal solution can only use $\wmax$ items among these bases, any $k$-subset in the future must contain at least one of the remaining items in these bases. There are only $\tO(1)$ primes $p_j$ that initially have a positive exponent $q_j$, and $q_j$ is bounded by $O(\log\wmax)$, so the algorithm will terminate in $\tO(1)$ rounds, generating only $\tO(k)$ bases in total.

Only the weights $w_i$ in the $\tO(k)$ bases can have more than $(2r+1)+r=3r+1$ items appearing in the optimal solution. Summing up for all batches and all values of $r$, we obtain our $\tO(\wmax)$ bound for the total number of items that we need to consider. Furthermore, we know the identity of these items. Easy to check that our algorithm runs in $\tO(n+\wmax)$ time.

\end{proof}


\paragraph{Main algorithm.} Using Theorem~\ref{thm:item_reduction}, we can reduce the bounded knapsack problem to a 0-1 knapsack problem (of the form \DK, which may contain negative items) with only $n=\tO(\wmax)$ items. One can check that our $\tO(n^{3/2}\wmax)$-time randomized algorithm for 0-1 knapsack in Theorem~\ref{thm:01_knapsack_random} also works with the presence of negative items, so we obtain the following result for bounded knapsack:
\begin{theorem}\label{thm:bounded_knapsack_wmax}
There is an $\tO(n+\wmax^{5/2})$-time randomized algorithm for bounded knapsack that is correct w.h.p.
\end{theorem}

\noindent By known techniques~\cite{DBLP:conf/icalp/0001RW21}, we can replace $\wmax$ with $\pmax$ in the running time of Theorem~\ref{thm:bounded_knapsack_wmax}.
\begin{theorem}\label{thm:bounded_knapsack_pmax}
There is an $\tO(n+\pmax^{5/2})$-time randomized algorithm for bounded knapsack that is correct w.h.p.
\end{theorem}

We emphasize that although the correctness proof for our algorithm is nontrivial, the algorithm itself is still simple.


\paragraph{Acknowledgement.} We thank Ce Jin for helpful discussion.

\bibliographystyle{plain}
\bibliography{references}
\end{document}